\documentclass[aps,prl,twocolumn,superscriptaddress,longbibliography,nofootinbib,secnumarabic]{revtex4-2}
\usepackage[T1]{fontenc}
\usepackage[utf8]{inputenc}
\usepackage[english]{babel}   
\usepackage{lmodern}
\usepackage{amsmath,amssymb,amsthm,bm,bbm}
\usepackage{siunitx}
\usepackage{xspace}
\usepackage{graphicx}
\usepackage{hyperref}
\usepackage{braket}
\usepackage{physics}        
\usepackage{enumitem}       
\usepackage{times}
\usepackage{siunitx}  
\usepackage{listings}
\usepackage{xcolor} 
\usepackage{mathtools }
\usepackage{tabularx}
\usepackage{tabularray}
\usepackage{threeparttable}
\usepackage{orcidlink} 
\usepackage{tcolorbox}
\usepackage{booktabs}

\hypersetup{
  colorlinks=true,
  linkcolor=magenta,
  citecolor=magenta,
  urlcolor=magenta
}

\newtcolorbox{draftrefs}{
  colback=gray!10,
  colframe=gray!50,
  title={References (for this section --- to verify later)},
  fonttitle=\bfseries
}

\renewcommand{\dd}{\mathrm{d}}

\begin{document}

\title{Velocity-Controlled Directional Readout of Single Photons}

\author{Mohamed Hatifi\,\orcidlink{0009-0005-3368-2751}}
\affiliation{Aix Marseille Univ, CNRS, Centrale M\'editerran\'ee, Institut Fresnel, Marseille, France}
\email{hatifi@fresnel.fr}

\begin{abstract}
Photodetection is usually treated in the frame in which the detector is at rest relative to the optical apparatus. We show that uniform motion of an electric Glauber detector changes the single-click POVM realized on two counterpropagating single-photon modes. Motion Doppler-shifts the alternatives in the detector frame; finite bandwidth then converts propagation direction into a detection bias without decohering the photon. For a Lorentzian response near one Doppler branch, the readout crosses from phase-sensitive to direction-sensitive with a quality-factor-enhanced onset. Finite-time integration adds Doppler-beat visibility loss, separating passive covariance from measurement change.
\end{abstract}

\maketitle

\paragraph{Introduction.—}
Photodetection is the operational step through which an optical quantum state becomes a measurement record. In the standard Glauber formulation, the detector is effectively at rest with respect to the optical apparatus, and an electric-dipole response probes the positive-frequency electric field through a normally ordered correlation function \cite{glauber1963,kelley1964,mandel1995,loudon2000}. This familiar setting masks a simple relativistic question: if the detector moves relative to the optical modes, what field amplitude does it measure? The underlying relativistic transformations are standard. Electric and magnetic fields mix under Lorentz transformations, and counterpropagating waves are Doppler shifted in opposite directions in the detector frame \cite{jackson2009}. Related issues arise in relativistic quantum information, moving-detector models, and generalized photodetection schemes with electric and magnetic response channels \cite{birrell1982,dragan2013,hawton2013,tanimura2014,hatifi2026b}. These works establish the importance of the detector frame and the covariance of field descriptions. What remains less explicit is the corresponding measurement-theoretic statement: for a fixed laboratory single-photon state, changing the detector velocity changes the positive-operator-valued measure (POVM) implemented by a realistic detector with finite spectral response \cite{davies1970,polo-gomez2022}.
\\
\indent The distinction is operational rather than merely descriptive: a passive Lorentz transformation of the complete experiment cannot change physical probabilities. By contrast, changing the detector velocity relative to a fixed laboratory state changes the rest-frame field sampled by the detector. In the broadband limit, this mainly produces a weak, kinematic electric--magnetic admixture. For a spectrally selective detector, however, the two propagation alternatives sample different points of the rest-frame response function \cite{clerk2010a,vanenk2017,propp2019}. Motion can therefore make a detection event direction-selective even when the incident single-photon state remains coherent. Here, we show that a moving electric Glauber detector realizes a velocity-dependent, frequency-selective photodetection POVM for a counterpropagating single photon. The detector velocity fixes both the relative weights of the two propagation alternatives and the proper-time Doppler beat between them. We derive the resulting visibility--bias relation for the instantaneous detection probability and show that, when a Lorentzian detector is operated near one Doppler branch, the crossover from phase-sensitive to direction-sensitive readout is set by the Doppler splitting in units of the detector linewidth. Thus, detector motion acquires a direct measurement-theoretic meaning: it changes the measurement axis selected by the detector on the single-photon propagation qubit. The predicted signatures are a Doppler beat in the detection record, a controllable visibility--bias tradeoff, and a crossover from interference-sensitive to which-way readout set by detector bandwidth rather than by a relativistic velocity scale alone. The most accessible tests are likely in microwave, circuit-QED, optomechanical, or time-modulated photonic platforms, where narrowband detection and synthetic Doppler shifts can replace literal high-speed optical motion \cite{kippenberg2008a,romero2009,fang2012,sathyamoorthy2016,sounas2017,besse2018,yuan2018,harwood2025}. More broadly, the result shows that detector motion can select the quantum observable realized by photodetection, separating passive Lorentz covariance from a physical change of measurement.
\paragraph{Rest-frame Glauber detection.—}
Consider a detector moving at constant velocity \(v=\beta c\) along the \(x\) axis of the laboratory frame. We choose the origin such that the detector crosses \(x=0\) at \(t=0\). Its proper time \(\tau\) and laboratory coordinates are then related by
\begin{equation}
t=\gamma\tau,\qquad x=\gamma\beta c\,\tau,
\qquad
\gamma=(1-\beta^2)^{-1/2}.
\label{eq:worldline}
\end{equation}
We take the detector to be purely electric in its own rest frame. The lowest-order instantaneous count rate, per unit detector proper time, is governed by the positive-frequency electric field evaluated on the detector worldline \cite{glauber1963,kelley1964},
\begin{equation}
R(\tau)\propto
\left\langle
\widehat{E}'^{(-)}_y(\tau)
\widehat{E}'^{(+)}_y(\tau)
\right\rangle ,
\label{eq:glauber_rest}
\end{equation}
where the prime denotes the detector rest frame and the positive-frequency part is defined with respect to the detector proper time. For transverse laboratory fields \(E_y\) and \(F_z=cB_z\), the Lorentz transformation to the detector frame gives \cite{jackson2009}
\begin{equation}
\widehat{E}'^{(+)}_y
=
\gamma\left(
\widehat{E}^{(+)}_y
-
\beta \widehat{F}^{(+)}_z
\right).
\label{eq:lorentz_E}
\end{equation}
For the inertial monochromatic modes considered below, the Doppler-shifted frequencies remain positive for \(|\beta|<1\), so Eq.~\eqref{eq:lorentz_E} is equivalent to transforming the full field and then retaining its positive-frequency part along the worldline.  Thus, a detector that is purely electric in its own frame is described in the laboratory frame by a velocity-fixed electric--magnetic detection amplitude \cite{durnin1981,tanimura2014,hatifi2026b}. 

\noindent We now apply this rest-frame detection rule to two counterpropagating laboratory modes \cite{akhlaghi2015,hatifi2026} with the same frequency \(\omega=ck\), polarization along \(y\), and field normalization \(\mathcal E\). Their positive-frequency electric field and rescaled magnetic field are
\begin{align}
\widehat{E}^{(+)}_y(x,t)
&=
\mathcal E
\left(
\hat a_+ e^{ikx-i\omega t}
+
\hat a_- e^{-ikx-i\omega t}
\right),
\label{eq:E_lab}
\\
\widehat{F}^{(+)}_z(x,t)
&=
\mathcal E
\left(
\hat a_+ e^{ikx-i\omega t}
-
\hat a_- e^{-ikx-i\omega t}
\right),
\label{eq:F_lab}
\end{align}
where \(+\) and \(-\) label propagation along \(+x\) and \(-x\), respectively. The opposite signs in Eq.~\eqref{eq:F_lab} express the reversal of the magnetic field under reversal of the propagation direction. Evaluated on the detector worldline, the two phases are
\[
kx-\omega t=-\gamma(1-\beta)\omega\tau,
\qquad
-kx-\omega t=-\gamma(1+\beta)\omega\tau .
\]
\begin{figure}[t]
\centering
\includegraphics[width=\columnwidth]{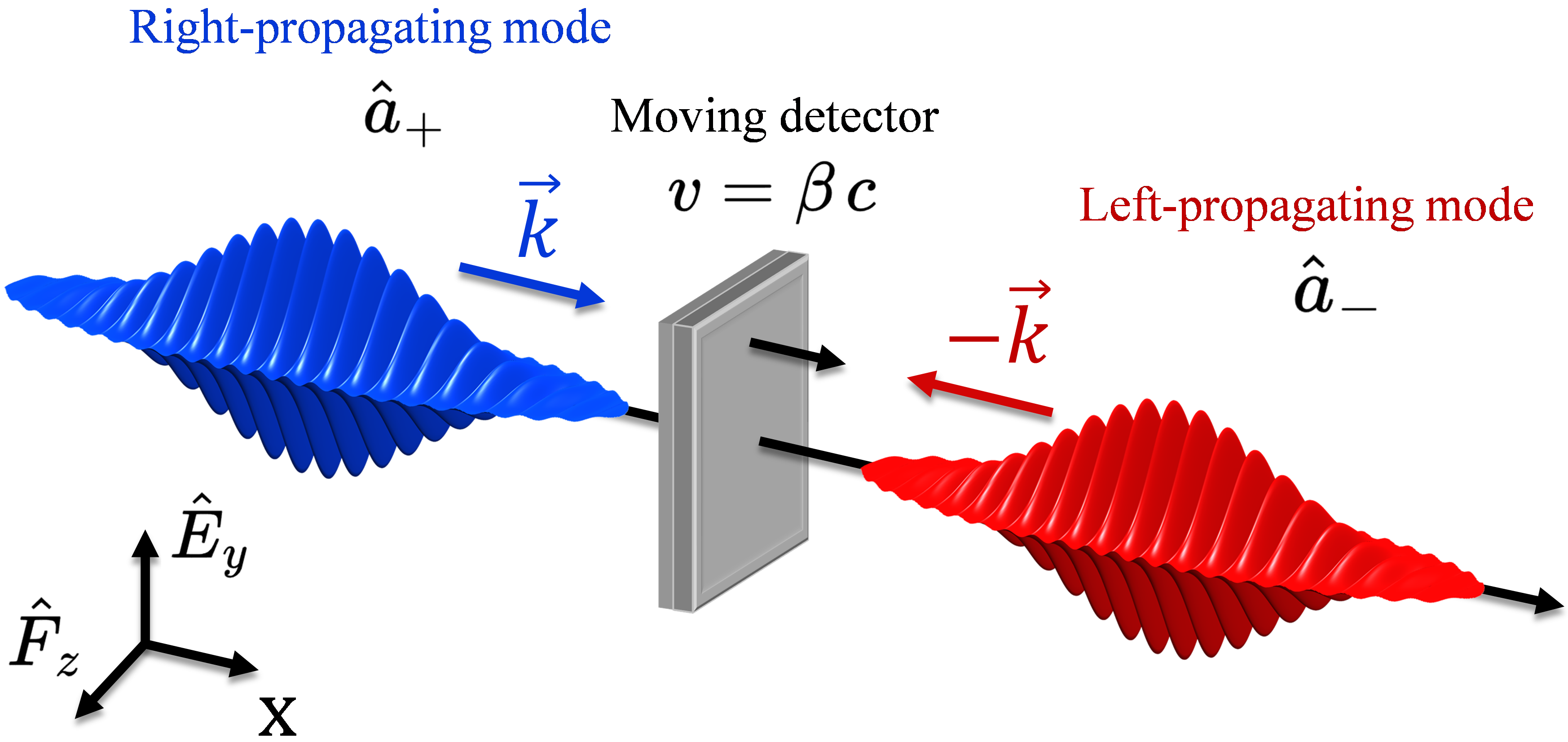}
\caption{
\textbf{Moving-detector geometry.} A detector moving along \(+x\) with velocity \(v=\beta c\) probes two counterpropagating single-photon modes of the same laboratory frequency \(\omega\). The right-propagating blue mode is annihilated by \(\hat a_+\) and carries wave vector \(+\mathbf{k}\), while the left-propagating red mode is annihilated by \(\hat a_-\) and carries wave vector \(-\mathbf{k}\). Although the detector is purely electric in its rest frame, its laboratory-frame response mixes electric and magnetic amplitudes and samples the two modes at the Doppler-shifted frequencies \(\Omega_\pm=\gamma(1\mp\beta)\omega\). This is the origin of the velocity-dependent photodetection POVM derived below.
}
\label{fig:setup}
\end{figure}
The detector therefore sees the two propagation alternatives at the Doppler-shifted frequencies
\begin{equation}
\Omega_+=\gamma(1-\beta)\omega,
\qquad
\Omega_-=\gamma(1+\beta)\omega,
\label{eq:doppler}
\end{equation}
with splitting $\Delta\Omega=\Omega_- - \Omega_+ = 2\gamma\beta\omega$. Combining Eqs.~\eqref{eq:lorentz_E}--\eqref{eq:F_lab}, the rest-frame electric field sampled by a broadband moving detector is
\begin{equation}
\widehat{E}'^{(+)}_y(\tau)
=
\gamma\mathcal E
\left[
(1-\beta)\hat a_+ e^{-i\Omega_+\tau}
+
(1+\beta)\hat a_- e^{-i\Omega_-\tau}
\right].
\label{eq:broad_operator}
\end{equation}
Equation~\eqref{eq:broad_operator} displays the two velocity-dependent structures inherited by the measurement operator. The amplitudes \(1\mp\beta\) encode Lorentz-induced electric--magnetic mixing, while the frequencies \(\Omega_\pm\) encode the Doppler splitting that a finite-bandwidth or finite-time detector can convert into operational distinguishability.

\paragraph{Velocity-dependent single-photon POVM.—}
We now include the detector's finite spectral response in its rest frame. For a linear stationary detector, the response acts as a convolution in proper time; for the monochromatic components considered here, this reduces to multiplication by a complex susceptibility \(\chi(\Omega)\) evaluated at the corresponding detector-frame frequency \cite{clerk2010a,vanenk2017,young2018,propp2019}. The detected positive-frequency amplitude is therefore
\begin{equation}
\widehat{\mathcal O}_\beta(\tau)
=
\mathcal E
\left[
g_+(\beta)\hat a_+ e^{-i\Omega_+\tau}
+
g_-(\beta)\hat a_- e^{-i\Omega_-\tau}
\right],
\label{eq:O_beta}
\end{equation}
with $g_\pm(\beta)=\gamma(1\mp\beta)\chi(\Omega_\pm)$.
Thus, the detector velocity enters the measurement operator in two distinct ways: through the Lorentz electric--magnetic weights \(1\mp\beta\), and through the detector-frame frequencies \(\Omega_\pm\) at which the rest-frame susceptibility is sampled. Let us consider now the single-photon state
\begin{equation}
\ket{\psi_\phi}
=
\frac{1}{\sqrt2}
\left(
\hat a_+^\dagger
+
e^{i\phi}\hat a_-^\dagger
\right)\ket{0},
\label{eq:single_photon}
\end{equation}
where \(\phi\) is the relative phase between the counterpropagating components. The corresponding single-click effect density is
\(\widehat{\Pi}_\beta(\tau)=
\widehat{\mathcal O}_\beta^\dagger(\tau)
\widehat{\mathcal O}_\beta(\tau)\), so the proper-time count rate for \(\ket{\psi_\phi}\) is \cite{srinivas1981,vanenk2017,helstrom2010}
\begin{align}
P_\beta(\tau,\phi)
&=
\mel{\psi_\phi}
{\widehat{\Pi}_\beta(\tau)}
{\psi_\phi}
\nonumber\\
&=
\frac{|\mathcal E|^2}{2}
\left[
|g_+|^2+|g_-|^2
+
2\operatorname{Re}
\left(
g_+^*g_-
e^{-i(\Delta\Omega\tau-\phi)}
\right)
\right].
\label{eq:P_beta}
\end{align}
\begin{figure}[t]
\centering
\includegraphics[width=\columnwidth]{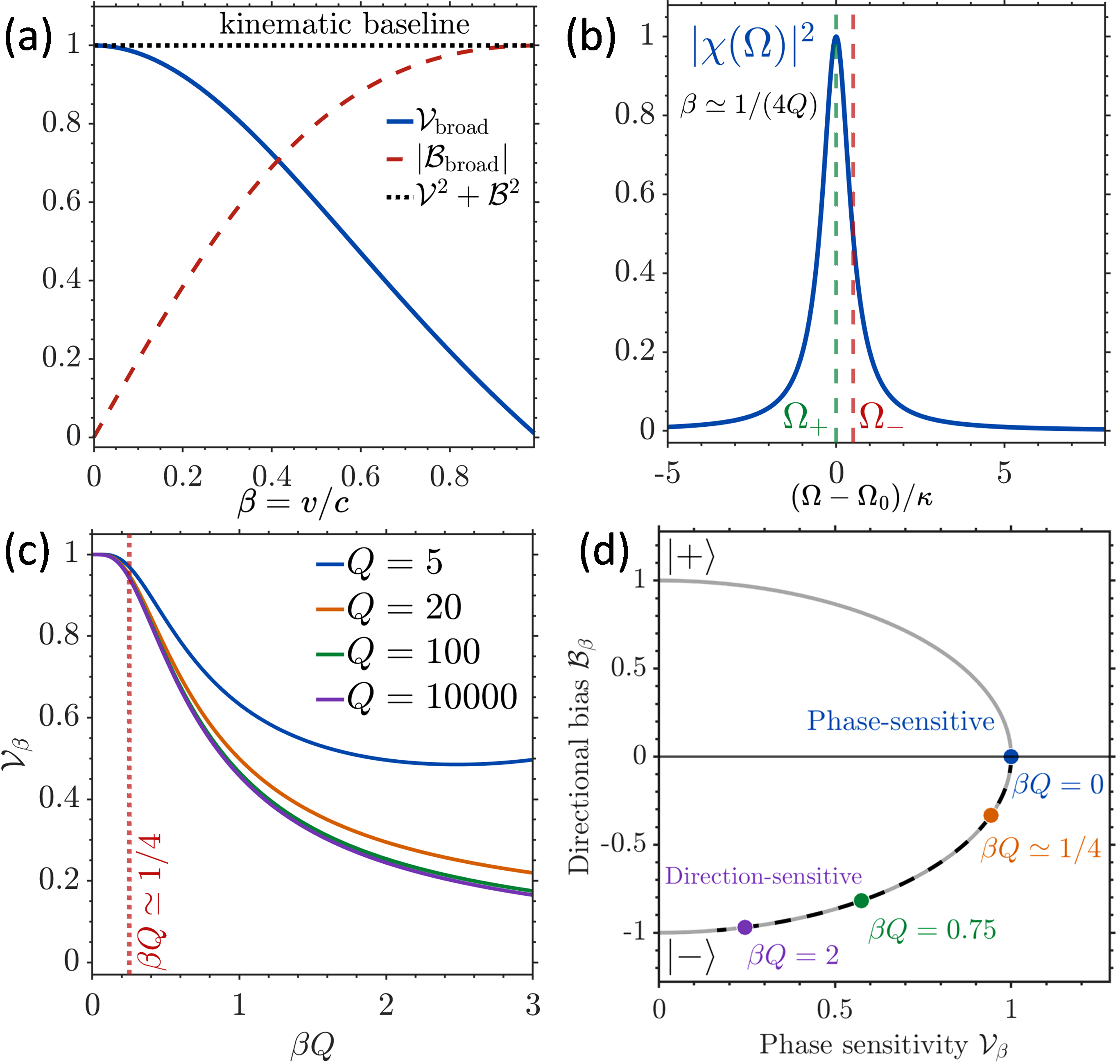}
\caption{
\textbf{Moving-detector photodetection and Doppler-enhanced directional readout.}
\textit{(a) Broadband limit.} A detector that is purely electric in its rest frame realizes, in the laboratory frame, a velocity-fixed electric--magnetic readout. The visibility \(\mathcal V_{\rm broad}\) decreases as the propagation bias \(|\mathcal B_{\rm broad}|\) increases, while the ideal click effect satisfies \(\mathcal V^2+\mathcal B^2=1\).
\textit{(b) Narrowband mechanism.} A Lorentzian detector tuned near one Doppler-shifted component samples the two counterpropagating alternatives at different detector-frame frequencies, \(\Omega_+\) and \(\Omega_-\). Propagation selectivity appears once the Doppler splitting resolves the linewidth.
\textit{(c) \(Q\)-enhanced crossover.} For a Lorentzian detector operated near one Doppler branch, the visibility collapses as a function of the scaled velocity \(\beta Q\), with the onset near \(\beta Q\simeq1/4\).
\textit{(d) Qubit-analyzer interpretation.} In the one-photon subspace \(\{\ket{+},\ket{-}\}\), the normalized click effect defines a Bloch-vector analyzer. The meridian projection shows the motion-induced displacement from an equatorial, phase-sensitive readout \((\mathcal V_\beta\simeq1,\mathcal B_\beta\simeq0)\) toward a polar, direction-sensitive readout \((|\mathcal B_\beta|\simeq1)\) as \(\beta Q\) increases.
}
\label{fig:moving_detector_summary}
\end{figure}
The first two terms are the direction-dependent click weights, while the last term is the phase-sensitive interference contribution. The phase of \(g_+^*g_-\) shifts the fringe phase, whereas its modulus fixes the fringe contrast. Because the two alternatives have different detector-frame frequencies, the interference term oscillates in proper time at the Doppler beat frequency \(\Delta\Omega\). The ideal two-mode detection effect is characterized by an instantaneous fringe visibility
\begin{equation}
\mathcal V_\beta
=
\frac{2|g_+g_-|}
{|g_+|^2+|g_-|^2},
\label{eq:visibility_general}
\end{equation}
and by the signed directional bias
\begin{equation}
\mathcal B_\beta
=
\frac{|g_+|^2-|g_-|^2}
{|g_+|^2+|g_-|^2}.
\label{eq:bias_general}
\end{equation}
Equivalently, after normalization in the one-photon subspace
\(\{\ket{+},\ket{-}\}\), the click effect can be written as a qubit analyzer,
\begin{equation}
\widehat{\pi}_\beta(\tau)
=
\frac12
\left[
\mathbb I+
\mathbf n_\beta(\tau)\cdot\boldsymbol{\sigma}
\right],
\label{eq:bloch_povm}
\end{equation}
with $\mathbf n_\beta(\tau)=\left(\mathcal V_\beta\cos\Theta_\beta(\tau),\mathcal V_\beta\sin\Theta_\beta(\tau),\mathcal B_\beta\right)$
and $\Theta_\beta(\tau)=\Delta\Omega\tau-\arg(g_+^*g_-)$. The detector velocity therefore selects the measurement axis of the propagation qubit: spectral selectivity shifts the click effect from an equatorial, phase-sensitive analyzer toward a polar, direction-sensitive analyzer. The sign of \(\mathcal B_\beta\) indicates which propagation direction is preferentially sampled, while \(|\mathcal B_\beta|\) is the operational which-way bias of the click event. For this ideal single-channel detection effect, $\mathcal V_\beta^2+\mathcal B_\beta^2=1$ \cite{wootters1979,greenberger1988,jaeger1995,englert1996,bera2015}. 

\noindent This relation is not due to a loss of coherence in the incident state \(\ket{\psi_\phi}\). It reflects the detector-selected POVM: as the detection effect becomes more biased toward one propagation direction, it becomes less sensitive to the relative phase between the two components. In the broadband limit, \(\chi(\Omega_+)=\chi(\Omega_-)\), the spectral response drops out. Equations~\eqref{eq:visibility_general} and \eqref{eq:bias_general} then give
\begin{equation}
\mathcal V_{\rm broad}(\beta)
=
\frac{1-\beta^2}{1+\beta^2},
\qquad
\mathcal B_{\rm broad}(\beta)
=
-\frac{2\beta}{1+\beta^2},
\label{eq:broad_VB}
\end{equation}
with the present convention that positive \(\beta\) denotes motion along \(+x\). This broadband limit is the covariant baseline: a moving electric detector realizes a velocity-fixed electric--magnetic readout \cite{tanimura2014,hatifi2026b}, but the visibility reduction is only quadratic in \(\beta\). The stronger crossover requires a spectrally selective detector response.
\\ 
\paragraph{\(Q\)-enhanced Doppler which-way transition.—}
The broadband result above is mostly kinematic. A parametrically stronger effect appears when the detector is spectrally selective. We model the detector response in its rest frame by the Lorentzian susceptibility \cite{gardiner1985c,gardiner2010a,walls2008,clerk2010a}
\begin{equation}
\chi(\Omega)
=
\frac{\chi_0}{\kappa/2-i(\Omega-\Omega_0)},
\label{eq:lorentzian}
\end{equation}
where \(\Omega_0\) is the detector resonance and \(\kappa\) is the full width at half maximum of \(|\chi(\Omega)|^2\). For an arbitrary detuning of the detector resonance, one finds
\begin{equation}
\frac{|g_-|}{|g_+|}
=
\frac{1+\beta}{1-\beta}
\left[
\frac{(\kappa/2)^2+(\Omega_+-\Omega_0)^2}
     {(\kappa/2)^2+(\Omega_- -\Omega_0)^2}
\right]^{1/2}.
\label{eq:ratio_general_detuning}
\end{equation}
This expression separates the broadband Lorentz amplitude imbalance, given by the prefactor \((1+\beta)/(1-\beta)\), from the genuinely dispersive contribution due to the finite detector linewidth. It also shows that direction selectivity depends on the detector's operating point, not only on the velocity. A particularly transparent branch-selective operating point is obtained by placing the resonance on a single Doppler-shifted component. Suppose, for definiteness, that the detector is tuned to the \(+\) component for the chosen velocity, $\Omega_0=\Omega_+$. The \(+\) component is then resonant with the detector, while the \(-\) component is detuned by the Doppler splitting $\Delta\Omega=\Omega_- - \Omega_+=2\gamma\beta\omega$. Equation~\eqref{eq:ratio_general_detuning} then reduces to
\begin{equation}
\frac{|g_-|}{|g_+|}
=
\frac{1+\beta}{1-\beta}
\frac{1}
{\sqrt{1+\left(4\gamma\beta\omega/\kappa\right)^2}} .
\label{eq:ratio}
\end{equation}
The first factor is the broadband Lorentz amplitude imbalance. The second factor is the dispersive suppression of the off-resonant propagation component. The onset of spectral direction selectivity occurs when the Doppler detuning reaches the half-width of the response, $\Delta\Omega=2\gamma\beta\omega \sim \frac{\kappa}{2}$. Equivalently, with \(Q\simeq\omega/\kappa\), the onset velocity satisfies $\beta_c\sim \frac{1}{4Q}$ in the nonrelativistic regime. This is the central enhancement mechanism: the detector linewidth promotes the small kinematic parameter \(\beta\) into the spectrally resolved control parameter \(\beta Q\). A narrowband detector operated near one Doppler branch can therefore convert a small Doppler shift into a large imbalance between the two detected propagation alternatives. The corresponding visibility and directional bias follow directly from the general expressions above. Writing $r(\beta)=|g_-|/|g_+|$, one finds
\begin{equation}
\mathcal V_\beta
=
\frac{2r}{1+r^2},
\qquad
|\mathcal B_\beta|
=
\frac{|1-r^2|}{1+r^2}.
\label{eq:Vr}
\end{equation}
Thus, the detector crosses over from phase-sensitive readout, \(r\simeq1\), to direction-sensitive readout, \(r\ll1\) or \(r\gg1\), as the Doppler splitting resolves the two propagation components within the detector response. This separates the present effect from the broadband Lorentz-mixing baseline of Eq.~\eqref{eq:broad_VB}: the visibility loss is now controlled by the spectrally enhanced parameter \(\beta Q\), rather than by the broadband kinematic scale \(\beta^2\).

\paragraph{Finite-time detection.—}
The preceding visibility is an instantaneous proper-time visibility. A physical count record, however, is often accumulated over a finite time bin or acquisition window \cite{srinivas1981,vanenk2017}. Here \(T\) denotes this external integration window; it is distinct from the internal response time associated with the susceptibility \(\chi(\Omega)\). If the signal is integrated over a proper-time interval \(T\), the interference term in Eq.~\eqref{eq:P_beta} is averaged over the Doppler beat. For a normalized rectangular gate, this average gives
\begin{equation}
\frac{1}{T}\int_0^T \dd\tau\, e^{-i\Delta\Omega\tau}
=
e^{-i\Delta\Omega T/2}
\operatorname{sinc}\left(\frac{\Delta\Omega T}{2}\right),
\label{eq:gate_integral}
\end{equation}
where \(\operatorname{sinc}x=\sin x/x\). The phase factor in Eq.~\eqref{eq:gate_integral} only shifts the observed fringe phase. The observed visibility is thus
\begin{equation}
\mathcal V_{\mathrm{obs}}(\beta)
=
\mathcal V_\beta
\left|
\operatorname{sinc}\left(\frac{\Delta\Omega T}{2}\right)
\right|
=
\mathcal V_\beta
\left|
\operatorname{sinc}\left(\gamma\beta\omega T\right)
\right|.
\label{eq:finite_time_visibility}
\end{equation}
Temporal coarse-graining, therefore, shrinks the transverse component of the detector-selected qubit analyzer while leaving the directional bias unchanged. For the rectangular gate,
\begin{equation}
\mathcal V_{\mathrm{obs}}^2+\mathcal B_\beta^2
=
\mathcal V_\beta^2
\left|
\operatorname{sinc}\left(\gamma\beta\omega T\right)
\right|^2
+
\mathcal B_\beta^2
\leq 1 .
\label{eq:finite_time_unsharp}
\end{equation}
Thus, finite integration converts the instantaneous rank-one click effect into an unsharp measurement of the propagation qubit. This gives a second route to motion-induced loss of observed interference. Even when the detector is not spectrally selective, the two propagation components beat at different detector-frame frequencies. If the count record is integrated over a window long enough that \(\gamma\beta\omega T\gtrsim1\), the phase-sensitive term is averaged out. The photon state remains coherent; what is lost is the operational visibility of that coherence when the record is integrated without conditioning on the Doppler-beat phase. In this sense, finite-time detection converts motion-induced frequency splitting into an apparent reduction of fringe contrast.
\begin{figure}[t]
\centering
\includegraphics[width=.935\columnwidth]{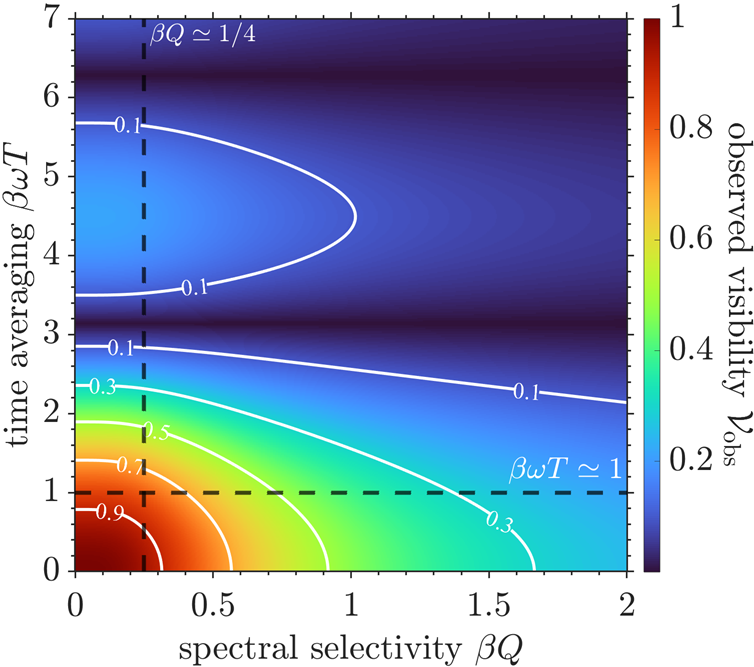}
\caption{
\textbf{Observed visibility in the presence of spectral selectivity and finite-time integration.}
The color map shows the visibility \(\mathcal V_{\mathrm{obs}}\) as a function of two dimensionless control parameters: the spectral-selectivity parameter \(\beta Q\), proportional to the Doppler splitting in linewidth units, and the time-averaging parameter \(\beta\omega T\), which measures the accumulated Doppler-beat phase during the detection window. The vertical dashed line marks the narrowband onset \(\beta Q\simeq1/4\) for a branch-tuned Lorentzian response, where the detector begins to resolve the two propagation alternatives spectrally. The horizontal dashed line marks \(\beta\omega T\simeq1\), where finite-time integration begins to wash out the Doppler beat. The contours show that visibility can be lost either by frequency-selective direction readout or by temporal averaging of an otherwise coherent interference signal.
}
\label{fig:visibility_map}
\end{figure}
\paragraph{Discussion and implementation routes.—} 
The results have two complementary interpretations. In the laboratory frame, a detector that is purely electric in its own rest frame realizes a velocity-dependent electric--magnetic detection operator. In the detector frame, the same physics appears as a Doppler splitting of the two propagation alternatives. The first viewpoint connects the result to generalized electric--magnetic photodetection \cite{tanimura2014,hatifi2026b}, while the second identifies the enhancement mechanism: finite detector bandwidth converts a small detector-frame frequency splitting into an operational directional bias. This distinction separates passive covariance from operational measurement change. A Lorentz transformation of the whole experiment only changes the description and leaves detection probabilities invariant. Here, the laboratory single-photon state is kept fixed, while the detector velocity is physically varied. The two propagation components then acquire different detector-frame frequencies and sample different points of the rest-frame response, so the implemented single-click POVM changes. 

\noindent The incident photon state need not lose coherence; rather, the detector-selected effect becomes less phase-sensitive and more direction-selective. Direct optical implementations with mechanically moving detectors are challenging because they require narrowband detection, controlled motion, and time- or phase-resolved readout at the relevant Doppler beat scale. More accessible routes are likely to be synthetic. The required elements are a two-directional mode basis, a detector response with controllable linewidth, and a physical or synthetic mechanism that shifts the two alternatives in opposite directions relative to the detector response. In such implementations, the literal Doppler splitting $\Delta\Omega=2\gamma\beta\omega$ is replaced by an engineered splitting \(\Delta\Omega_{\rm eff}\), and the relevant selectivity parameter becomes \(\Delta\Omega_{\rm eff}/\kappa\). Microwave photons, superconducting circuits, cavity or waveguide detectors with tunable narrowband response, optomechanical motion, electro-optic modulation, and time-modulated photonic structures provide natural settings in which this condition may be engineered \cite{blais2004,wallraff2004,romero2009,fang2012,aspelmeyer2014a,sounas2017,besse2018,yuan2018,holzman2019,galiffi2022,harwood2025a}.  Experimentally, the effect would appear as a detector-frame beat in the count record, a visibility--bias tradeoff for the ideal two-mode click effect, and, for a detector operated near one Doppler branch, a crossover from phase-sensitive to direction-sensitive readout controlled by the ratio of the induced splitting to the detector linewidth. 
\\

\noindent The same incident single-photon state can therefore be read either through its phase coherence or through a directional bias, depending on the click effect selected by the detector motion.

\paragraph{Conclusion.—}
We have shown that Glauber photodetection, when formulated in the detector rest frame, acquires a velocity-dependent operational meaning for a detector moving through a two-mode single-photon field. In the broadband limit, the effect reduces to the Lorentz-covariant mixing of electric and magnetic field amplitudes, yielding a constrained electric--magnetic detection operator in the laboratory frame. With a finite spectral response, however, motion does more: it correlates the propagation direction with the detector-frame frequency, so the two components sample different points of the detector response, and the implemented single-click POVM changes.  For a Lorentzian response operated near one Doppler branch, this produces a \(Q\)-enhanced crossover from phase-sensitive to direction-sensitive readout, while finite-time integration further suppresses the observed visibility through Doppler-beat averaging. Thus, detector motion provides a concrete control parameter for quantum photodetection: the detector trajectory and response function jointly select the observable realized on the photon, separating passive Lorentz covariance from a physical change of measurement.

\paragraph{Acknowledgements—}
The author would like to thank Brian Stout, Thomas Durt, and Branko Kolaric for stimulating discussions. This work was supported by the EU: the EIC Pathfinder Challenges 2022 call through the Research Grant 101115149 (project ARTEMIS)


\begin{thebibliography}{43}%
\makeatletter
\providecommand \@ifxundefined [1]{%
 \@ifx{#1\undefined}
}%
\providecommand \@ifnum [1]{%
 \ifnum #1\expandafter \@firstoftwo
 \else \expandafter \@secondoftwo
 \fi
}%
\providecommand \@ifx [1]{%
 \ifx #1\expandafter \@firstoftwo
 \else \expandafter \@secondoftwo
 \fi
}%
\providecommand \natexlab [1]{#1}%
\providecommand \enquote  [1]{``#1''}%
\providecommand \bibnamefont  [1]{#1}%
\providecommand \bibfnamefont [1]{#1}%
\providecommand \citenamefont [1]{#1}%
\providecommand \href@noop [0]{\@secondoftwo}%
\providecommand \href [0]{\begingroup \@sanitize@url \@href}%
\providecommand \@href[1]{\@@startlink{#1}\@@href}%
\providecommand \@@href[1]{\endgroup#1\@@endlink}%
\providecommand \@sanitize@url [0]{\catcode `\\12\catcode `\$12\catcode
  `\&12\catcode `\#12\catcode `\^12\catcode `\_12\catcode `\%12\relax}%
\providecommand \@@startlink[1]{}%
\providecommand \@@endlink[0]{}%
\providecommand \url  [0]{\begingroup\@sanitize@url \@url }%
\providecommand \@url [1]{\endgroup\@href {#1}{\urlprefix }}%
\providecommand \urlprefix  [0]{URL }%
\providecommand \Eprint [0]{\href }%
\providecommand \doibase [0]{https://doi.org/}%
\providecommand \selectlanguage [0]{\@gobble}%
\providecommand \bibinfo  [0]{\@secondoftwo}%
\providecommand \bibfield  [0]{\@secondoftwo}%
\providecommand \translation [1]{[#1]}%
\providecommand \BibitemOpen [0]{}%
\providecommand \bibitemStop [0]{}%
\providecommand \bibitemNoStop [0]{.\EOS\space}%
\providecommand \EOS [0]{\spacefactor3000\relax}%
\providecommand \BibitemShut  [1]{\csname bibitem#1\endcsname}%
\let\auto@bib@innerbib\@empty
\bibitem [{\citenamefont {Glauber}(1963)}]{glauber1963}%
  \BibitemOpen
  \bibfield  {author} {\bibinfo {author} {\bibfnamefont {R.~J.}\ \bibnamefont
  {Glauber}},\ }\href {https://doi.org/10.1103/PhysRev.130.2529} {\bibfield
  {journal} {\bibinfo  {journal} {Physical Review}\ }\textbf {\bibinfo {volume}
  {130}},\ \bibinfo {pages} {2529} (\bibinfo {year} {1963})}\BibitemShut
  {NoStop}%
\bibitem [{\citenamefont {Kelley}\ and\ \citenamefont
  {Kleiner}(1964)}]{kelley1964}%
  \BibitemOpen
  \bibfield  {author} {\bibinfo {author} {\bibfnamefont {P.~L.}\ \bibnamefont
  {Kelley}}\ and\ \bibinfo {author} {\bibfnamefont {W.~H.}\ \bibnamefont
  {Kleiner}},\ }\href {https://doi.org/10.1103/PhysRev.136.A316} {\bibfield
  {journal} {\bibinfo  {journal} {Physical Review}\ }\textbf {\bibinfo {volume}
  {136}},\ \bibinfo {pages} {A316} (\bibinfo {year} {1964})}\BibitemShut
  {NoStop}%
\bibitem [{\citenamefont {Mandel}\ and\ \citenamefont
  {Wolf}(1995)}]{mandel1995}%
  \BibitemOpen
  \bibfield  {author} {\bibinfo {author} {\bibfnamefont {L.}~\bibnamefont
  {Mandel}}\ and\ \bibinfo {author} {\bibfnamefont {E.}~\bibnamefont {Wolf}},\
  }\href@noop {} {\emph {\bibinfo {title} {Optical Coherence and Quantum
  Optics}}}\ (\bibinfo  {publisher} {Cambridge University Press},\ \bibinfo
  {address} {Cambridge},\ \bibinfo {year} {1995})\BibitemShut {NoStop}%
\bibitem [{\citenamefont {Loudon}(2000)}]{loudon2000}%
  \BibitemOpen
  \bibfield  {author} {\bibinfo {author} {\bibfnamefont {R.}~\bibnamefont
  {Loudon}},\ }\href@noop {} {\emph {\bibinfo {title} {The {{Quantum Theory}}
  of {{Light}}}}},\ \bibinfo {edition} {3rd}\ ed.\ (\bibinfo  {publisher}
  {Oxford University Press},\ \bibinfo {address} {Oxford},\ \bibinfo {year}
  {2000})\BibitemShut {NoStop}%
\bibitem [{\citenamefont {Jackson}(2009)}]{jackson2009}%
  \BibitemOpen
  \bibfield  {author} {\bibinfo {author} {\bibfnamefont {J.~D.}\ \bibnamefont
  {Jackson}},\ }\href@noop {} {\emph {\bibinfo {title} {Classical
  Electrodynamics}}},\ \bibinfo {edition} {3rd}\ ed.\ (\bibinfo  {publisher}
  {Wiley},\ \bibinfo {address} {Hoboken, NY},\ \bibinfo {year}
  {2009})\BibitemShut {NoStop}%
\bibitem [{\citenamefont {Birrell}\ and\ \citenamefont
  {Davies}(1982)}]{birrell1982}%
  \BibitemOpen
  \bibfield  {author} {\bibinfo {author} {\bibfnamefont {N.~D.}\ \bibnamefont
  {Birrell}}\ and\ \bibinfo {author} {\bibfnamefont {P.~C.~W.}\ \bibnamefont
  {Davies}},\ }\href {https://doi.org/10.1017/CBO9780511622632} {\emph
  {\bibinfo {title} {Quantum {{Fields}} in {{Curved Space}}}}},\ \bibinfo
  {edition} {1st}\ ed.\ (\bibinfo  {publisher} {Cambridge University Press},\
  \bibinfo {year} {1982})\BibitemShut {NoStop}%
\bibitem [{\citenamefont {Dragan}\ \emph {et~al.}(2013)\citenamefont {Dragan},
  \citenamefont {Doukas}, \citenamefont {{Mart{\'i}n-Mart{\'i}nez}},\ and\
  \citenamefont {Bruschi}}]{dragan2013}%
  \BibitemOpen
  \bibfield  {author} {\bibinfo {author} {\bibfnamefont {A.}~\bibnamefont
  {Dragan}}, \bibinfo {author} {\bibfnamefont {J.}~\bibnamefont {Doukas}},
  \bibinfo {author} {\bibfnamefont {E.}~\bibnamefont
  {{Mart{\'i}n-Mart{\'i}nez}}},\ and\ \bibinfo {author} {\bibfnamefont {D.~E.}\
  \bibnamefont {Bruschi}},\ }\href
  {https://doi.org/10.1088/0264-9381/30/23/235006} {\bibfield  {journal}
  {\bibinfo  {journal} {Classical and Quantum Gravity}\ }\textbf {\bibinfo
  {volume} {30}},\ \bibinfo {pages} {235006} (\bibinfo {year}
  {2013})}\BibitemShut {NoStop}%
\bibitem [{\citenamefont {Hawton}(2013)}]{hawton2013}%
  \BibitemOpen
  \bibfield  {author} {\bibinfo {author} {\bibfnamefont {M.}~\bibnamefont
  {Hawton}},\ }\href {https://doi.org/10.1103/PhysRevA.87.042116} {\bibfield
  {journal} {\bibinfo  {journal} {Physical Review A}\ }\textbf {\bibinfo
  {volume} {87}},\ \bibinfo {pages} {042116} (\bibinfo {year}
  {2013})}\BibitemShut {NoStop}%
\bibitem [{\citenamefont {Tanimura}(2014)}]{tanimura2014}%
  \BibitemOpen
  \bibfield  {author} {\bibinfo {author} {\bibfnamefont {S.}~\bibnamefont
  {Tanimura}},\ }\href {https://doi.org/10.1088/0031-8949/2014/T160/014039}
  {\bibfield  {journal} {\bibinfo  {journal} {Physica Scripta}\ }\textbf
  {\bibinfo {volume} {T160}},\ \bibinfo {pages} {014039} (\bibinfo {year}
  {2014})}\BibitemShut {NoStop}%
\bibitem [{\citenamefont {Hatifi}\ and\ \citenamefont
  {Stout}(2026)}]{hatifi2026b}%
  \BibitemOpen
  \bibfield  {author} {\bibinfo {author} {\bibfnamefont {M.}~\bibnamefont
  {Hatifi}}\ and\ \bibinfo {author} {\bibfnamefont {B.}~\bibnamefont {Stout}},\
  }\href {https://doi.org/10.48550/ARXIV.2605.16886} {\bibinfo {title} {Basis-
  and {{Channel-Selective Quantum Photodetection}}}} (\bibinfo {year}
  {2026})\BibitemShut {NoStop}%
\bibitem [{\citenamefont {Davies}\ and\ \citenamefont
  {Lewis}(1970)}]{davies1970}%
  \BibitemOpen
  \bibfield  {author} {\bibinfo {author} {\bibfnamefont {E.~B.}\ \bibnamefont
  {Davies}}\ and\ \bibinfo {author} {\bibfnamefont {J.~T.}\ \bibnamefont
  {Lewis}},\ }\href {https://doi.org/10.1007/BF01647093} {\bibfield  {journal}
  {\bibinfo  {journal} {Communications in Mathematical Physics}\ }\textbf
  {\bibinfo {volume} {17}},\ \bibinfo {pages} {239} (\bibinfo {year}
  {1970})}\BibitemShut {NoStop}%
\bibitem [{\citenamefont {{Polo-G{\'o}mez}}\ \emph {et~al.}(2022)\citenamefont
  {{Polo-G{\'o}mez}}, \citenamefont {Garay},\ and\ \citenamefont
  {{Mart{\'i}n-Mart{\'i}nez}}}]{polo-gomez2022}%
  \BibitemOpen
  \bibfield  {author} {\bibinfo {author} {\bibfnamefont {J.}~\bibnamefont
  {{Polo-G{\'o}mez}}}, \bibinfo {author} {\bibfnamefont {L.~J.}\ \bibnamefont
  {Garay}},\ and\ \bibinfo {author} {\bibfnamefont {E.}~\bibnamefont
  {{Mart{\'i}n-Mart{\'i}nez}}},\ }\href
  {https://doi.org/10.1103/PhysRevD.105.065003} {\bibfield  {journal} {\bibinfo
   {journal} {Physical Review D}\ }\textbf {\bibinfo {volume} {105}},\ \bibinfo
  {pages} {065003} (\bibinfo {year} {2022})}\BibitemShut {NoStop}%
\bibitem [{\citenamefont {Clerk}\ \emph {et~al.}(2010)\citenamefont {Clerk},
  \citenamefont {Devoret}, \citenamefont {Girvin}, \citenamefont {Marquardt},\
  and\ \citenamefont {Schoelkopf}}]{clerk2010a}%
  \BibitemOpen
  \bibfield  {author} {\bibinfo {author} {\bibfnamefont {A.~A.}\ \bibnamefont
  {Clerk}}, \bibinfo {author} {\bibfnamefont {M.~H.}\ \bibnamefont {Devoret}},
  \bibinfo {author} {\bibfnamefont {S.~M.}\ \bibnamefont {Girvin}}, \bibinfo
  {author} {\bibfnamefont {F.}~\bibnamefont {Marquardt}},\ and\ \bibinfo
  {author} {\bibfnamefont {R.~J.}\ \bibnamefont {Schoelkopf}},\ }\href
  {https://doi.org/10.1103/RevModPhys.82.1155} {\bibfield  {journal} {\bibinfo
  {journal} {Reviews of Modern Physics}\ }\textbf {\bibinfo {volume} {82}},\
  \bibinfo {pages} {1155} (\bibinfo {year} {2010})}\BibitemShut {NoStop}%
\bibitem [{\citenamefont {Van~Enk}(2017)}]{vanenk2017}%
  \BibitemOpen
  \bibfield  {author} {\bibinfo {author} {\bibfnamefont {S.~J.}\ \bibnamefont
  {Van~Enk}},\ }\href {https://doi.org/10.1103/PhysRevA.96.033834} {\bibfield
  {journal} {\bibinfo  {journal} {Physical Review A}\ }\textbf {\bibinfo
  {volume} {96}},\ \bibinfo {pages} {033834} (\bibinfo {year}
  {2017})}\BibitemShut {NoStop}%
\bibitem [{\citenamefont {Propp}\ and\ \citenamefont
  {Van~Enk}(2019)}]{propp2019}%
  \BibitemOpen
  \bibfield  {author} {\bibinfo {author} {\bibfnamefont {T.~B.}\ \bibnamefont
  {Propp}}\ and\ \bibinfo {author} {\bibfnamefont {S.~J.}\ \bibnamefont
  {Van~Enk}},\ }\href {https://doi.org/10.1103/PhysRevA.100.033836} {\bibfield
  {journal} {\bibinfo  {journal} {Physical Review A}\ }\textbf {\bibinfo
  {volume} {100}},\ \bibinfo {pages} {033836} (\bibinfo {year}
  {2019})}\BibitemShut {NoStop}%
\bibitem [{\citenamefont {Kippenberg}\ and\ \citenamefont
  {Vahala}(2008)}]{kippenberg2008a}%
  \BibitemOpen
  \bibfield  {author} {\bibinfo {author} {\bibfnamefont {T.~J.}\ \bibnamefont
  {Kippenberg}}\ and\ \bibinfo {author} {\bibfnamefont {K.~J.}\ \bibnamefont
  {Vahala}},\ }\href {https://doi.org/10.1126/science.1156032} {\bibfield
  {journal} {\bibinfo  {journal} {Science}\ }\textbf {\bibinfo {volume}
  {321}},\ \bibinfo {pages} {1172} (\bibinfo {year} {2008})}\BibitemShut
  {NoStop}%
\bibitem [{\citenamefont {Romero}\ \emph {et~al.}(2009)\citenamefont {Romero},
  \citenamefont {{Garc{\'i}a-Ripoll}},\ and\ \citenamefont
  {Solano}}]{romero2009}%
  \BibitemOpen
  \bibfield  {author} {\bibinfo {author} {\bibfnamefont {G.}~\bibnamefont
  {Romero}}, \bibinfo {author} {\bibfnamefont {J.~J.}\ \bibnamefont
  {{Garc{\'i}a-Ripoll}}},\ and\ \bibinfo {author} {\bibfnamefont
  {E.}~\bibnamefont {Solano}},\ }\href
  {https://doi.org/10.1103/PhysRevLett.102.173602} {\bibfield  {journal}
  {\bibinfo  {journal} {Physical Review Letters}\ }\textbf {\bibinfo {volume}
  {102}},\ \bibinfo {pages} {173602} (\bibinfo {year} {2009})}\BibitemShut
  {NoStop}%
\bibitem [{\citenamefont {Fang}\ \emph {et~al.}(2012)\citenamefont {Fang},
  \citenamefont {Yu},\ and\ \citenamefont {Fan}}]{fang2012}%
  \BibitemOpen
  \bibfield  {author} {\bibinfo {author} {\bibfnamefont {K.}~\bibnamefont
  {Fang}}, \bibinfo {author} {\bibfnamefont {Z.}~\bibnamefont {Yu}},\ and\
  \bibinfo {author} {\bibfnamefont {S.}~\bibnamefont {Fan}},\ }\href
  {https://doi.org/10.1038/nphoton.2012.236} {\bibfield  {journal} {\bibinfo
  {journal} {Nature Photonics}\ }\textbf {\bibinfo {volume} {6}},\ \bibinfo
  {pages} {782} (\bibinfo {year} {2012})}\BibitemShut {NoStop}%
\bibitem [{\citenamefont {Sathyamoorthy}\ \emph {et~al.}(2016)\citenamefont
  {Sathyamoorthy}, \citenamefont {Stace},\ and\ \citenamefont
  {Johansson}}]{sathyamoorthy2016}%
  \BibitemOpen
  \bibfield  {author} {\bibinfo {author} {\bibfnamefont {S.~R.}\ \bibnamefont
  {Sathyamoorthy}}, \bibinfo {author} {\bibfnamefont {T.~M.}\ \bibnamefont
  {Stace}},\ and\ \bibinfo {author} {\bibfnamefont {G.}~\bibnamefont
  {Johansson}},\ }\href {https://doi.org/10.1016/j.crhy.2016.07.010} {\bibfield
   {journal} {\bibinfo  {journal} {Comptes Rendus. Physique}\ }\textbf
  {\bibinfo {volume} {17}},\ \bibinfo {pages} {756} (\bibinfo {year}
  {2016})}\BibitemShut {NoStop}%
\bibitem [{\citenamefont {Sounas}\ and\ \citenamefont
  {Al{\`u}}(2017)}]{sounas2017}%
  \BibitemOpen
  \bibfield  {author} {\bibinfo {author} {\bibfnamefont {D.~L.}\ \bibnamefont
  {Sounas}}\ and\ \bibinfo {author} {\bibfnamefont {A.}~\bibnamefont
  {Al{\`u}}},\ }\href {https://doi.org/10.1038/s41566-017-0051-x} {\bibfield
  {journal} {\bibinfo  {journal} {Nature Photonics}\ }\textbf {\bibinfo
  {volume} {11}},\ \bibinfo {pages} {774} (\bibinfo {year} {2017})}\BibitemShut
  {NoStop}%
\bibitem [{\citenamefont {Besse}\ \emph {et~al.}(2018)\citenamefont {Besse},
  \citenamefont {Gasparinetti}, \citenamefont {Collodo}, \citenamefont
  {Walter}, \citenamefont {Kurpiers}, \citenamefont {Pechal}, \citenamefont
  {Eichler},\ and\ \citenamefont {Wallraff}}]{besse2018}%
  \BibitemOpen
  \bibfield  {author} {\bibinfo {author} {\bibfnamefont {J.-C.}\ \bibnamefont
  {Besse}}, \bibinfo {author} {\bibfnamefont {S.}~\bibnamefont {Gasparinetti}},
  \bibinfo {author} {\bibfnamefont {M.~C.}\ \bibnamefont {Collodo}}, \bibinfo
  {author} {\bibfnamefont {T.}~\bibnamefont {Walter}}, \bibinfo {author}
  {\bibfnamefont {P.}~\bibnamefont {Kurpiers}}, \bibinfo {author}
  {\bibfnamefont {M.}~\bibnamefont {Pechal}}, \bibinfo {author} {\bibfnamefont
  {C.}~\bibnamefont {Eichler}},\ and\ \bibinfo {author} {\bibfnamefont
  {A.}~\bibnamefont {Wallraff}},\ }\href
  {https://doi.org/10.1103/PhysRevX.8.021003} {\bibfield  {journal} {\bibinfo
  {journal} {Physical Review X}\ }\textbf {\bibinfo {volume} {8}},\ \bibinfo
  {pages} {021003} (\bibinfo {year} {2018})}\BibitemShut {NoStop}%
\bibitem [{\citenamefont {Yuan}\ \emph {et~al.}(2018)\citenamefont {Yuan},
  \citenamefont {Lin}, \citenamefont {Xiao},\ and\ \citenamefont
  {Fan}}]{yuan2018}%
  \BibitemOpen
  \bibfield  {author} {\bibinfo {author} {\bibfnamefont {L.}~\bibnamefont
  {Yuan}}, \bibinfo {author} {\bibfnamefont {Q.}~\bibnamefont {Lin}}, \bibinfo
  {author} {\bibfnamefont {M.}~\bibnamefont {Xiao}},\ and\ \bibinfo {author}
  {\bibfnamefont {S.}~\bibnamefont {Fan}},\ }\href
  {https://doi.org/10.1364/OPTICA.5.001396} {\bibfield  {journal} {\bibinfo
  {journal} {Optica}\ }\textbf {\bibinfo {volume} {5}},\ \bibinfo {pages}
  {1396} (\bibinfo {year} {2018})}\BibitemShut {NoStop}%
\bibitem [{\citenamefont {Harwood}\ \emph
  {et~al.}(2025{\natexlab{a}})\citenamefont {Harwood}, \citenamefont {Vezzoli},
  \citenamefont {Raziman}, \citenamefont {Hooper}, \citenamefont {Tirole},
  \citenamefont {Wu}, \citenamefont {Maier}, \citenamefont {Pendry},
  \citenamefont {Horsley},\ and\ \citenamefont {Sapienza}}]{harwood2025}%
  \BibitemOpen
  \bibfield  {author} {\bibinfo {author} {\bibfnamefont {A.~C.}\ \bibnamefont
  {Harwood}}, \bibinfo {author} {\bibfnamefont {S.}~\bibnamefont {Vezzoli}},
  \bibinfo {author} {\bibfnamefont {T.~V.}\ \bibnamefont {Raziman}}, \bibinfo
  {author} {\bibfnamefont {C.}~\bibnamefont {Hooper}}, \bibinfo {author}
  {\bibfnamefont {R.}~\bibnamefont {Tirole}}, \bibinfo {author} {\bibfnamefont
  {F.}~\bibnamefont {Wu}}, \bibinfo {author} {\bibfnamefont {S.~A.}\
  \bibnamefont {Maier}}, \bibinfo {author} {\bibfnamefont {J.~B.}\ \bibnamefont
  {Pendry}}, \bibinfo {author} {\bibfnamefont {S.~A.~R.}\ \bibnamefont
  {Horsley}},\ and\ \bibinfo {author} {\bibfnamefont {R.}~\bibnamefont
  {Sapienza}},\ }\href {https://doi.org/10.1038/s41467-025-60159-9} {\bibfield
  {journal} {\bibinfo  {journal} {Nature Communications}\ }\textbf {\bibinfo
  {volume} {16}},\ \bibinfo {pages} {5147} (\bibinfo {year}
  {2025}{\natexlab{a}})}\BibitemShut {NoStop}%
\bibitem [{\citenamefont {Durnin}\ \emph {et~al.}(1981)\citenamefont {Durnin},
  \citenamefont {Reece},\ and\ \citenamefont {Mandel}}]{durnin1981}%
  \BibitemOpen
  \bibfield  {author} {\bibinfo {author} {\bibfnamefont {J.}~\bibnamefont
  {Durnin}}, \bibinfo {author} {\bibfnamefont {C.}~\bibnamefont {Reece}},\ and\
  \bibinfo {author} {\bibfnamefont {L.}~\bibnamefont {Mandel}},\ }\href
  {https://doi.org/10.1364/JOSA.71.000115} {\bibfield  {journal} {\bibinfo
  {journal} {Journal of the Optical Society of America}\ }\textbf {\bibinfo
  {volume} {71}},\ \bibinfo {pages} {115} (\bibinfo {year} {1981})}\BibitemShut
  {NoStop}%
\bibitem [{\citenamefont {Akhlaghi}\ \emph {et~al.}(2015)\citenamefont
  {Akhlaghi}, \citenamefont {Schelew},\ and\ \citenamefont
  {Young}}]{akhlaghi2015}%
  \BibitemOpen
  \bibfield  {author} {\bibinfo {author} {\bibfnamefont {M.~K.}\ \bibnamefont
  {Akhlaghi}}, \bibinfo {author} {\bibfnamefont {E.}~\bibnamefont {Schelew}},\
  and\ \bibinfo {author} {\bibfnamefont {J.~F.}\ \bibnamefont {Young}},\ }\href
  {https://doi.org/10.1038/ncomms9233} {\bibfield  {journal} {\bibinfo
  {journal} {Nature Communications}\ }\textbf {\bibinfo {volume} {6}},\
  \bibinfo {pages} {8233} (\bibinfo {year} {2015})}\BibitemShut {NoStop}%
\bibitem [{\citenamefont {Hatifi}(2026)}]{hatifi2026}%
  \BibitemOpen
  \bibfield  {author} {\bibinfo {author} {\bibfnamefont {M.}~\bibnamefont
  {Hatifi}},\ }\href {https://doi.org/10.1103/mycc-r3bx} {\bibfield  {journal}
  {\bibinfo  {journal} {Physical Review A}\ }\textbf {\bibinfo {volume}
  {113}},\ \bibinfo {pages} {022204} (\bibinfo {year} {2026})}\BibitemShut
  {NoStop}%
\bibitem [{\citenamefont {Young}\ \emph {et~al.}(2018)\citenamefont {Young},
  \citenamefont {Sarovar},\ and\ \citenamefont {L{\'e}onard}}]{young2018}%
  \BibitemOpen
  \bibfield  {author} {\bibinfo {author} {\bibfnamefont {S.~M.}\ \bibnamefont
  {Young}}, \bibinfo {author} {\bibfnamefont {M.}~\bibnamefont {Sarovar}},\
  and\ \bibinfo {author} {\bibfnamefont {F.}~\bibnamefont {L{\'e}onard}},\
  }\href {https://doi.org/10.1103/PhysRevA.98.063835} {\bibfield  {journal}
  {\bibinfo  {journal} {Physical Review A}\ }\textbf {\bibinfo {volume} {98}},\
  \bibinfo {pages} {063835} (\bibinfo {year} {2018})}\BibitemShut {NoStop}%
\bibitem [{\citenamefont {Srinivas}\ and\ \citenamefont
  {Davies}(1981)}]{srinivas1981}%
  \BibitemOpen
  \bibfield  {author} {\bibinfo {author} {\bibfnamefont {M.}~\bibnamefont
  {Srinivas}}\ and\ \bibinfo {author} {\bibfnamefont {E.}~\bibnamefont
  {Davies}},\ }\href {https://doi.org/10.1080/713820643} {\bibfield  {journal}
  {\bibinfo  {journal} {Optica Acta: International Journal of Optics}\ }\textbf
  {\bibinfo {volume} {28}},\ \bibinfo {pages} {981} (\bibinfo {year}
  {1981})}\BibitemShut {NoStop}%
\bibitem [{\citenamefont {Helstrom}(2010)}]{helstrom2010}%
  \BibitemOpen
  \bibinfo {editor} {\bibfnamefont {C.~W.}\ \bibnamefont {Helstrom}},\ ed.,\
  \href@noop {} {\emph {\bibinfo {title} {Quantum Detection and Estimation
  Theory}}},\ \bibinfo {series} {Mathematics in Science and Engineering}\ No.\
  \bibinfo {number} {v. 123}\ (\bibinfo  {publisher} {Academic Press},\
  \bibinfo {address} {New York},\ \bibinfo {year} {2010})\BibitemShut {NoStop}%
\bibitem [{\citenamefont {Wootters}\ and\ \citenamefont
  {Zurek}(1979)}]{wootters1979}%
  \BibitemOpen
  \bibfield  {author} {\bibinfo {author} {\bibfnamefont {W.~K.}\ \bibnamefont
  {Wootters}}\ and\ \bibinfo {author} {\bibfnamefont {W.~H.}\ \bibnamefont
  {Zurek}},\ }\href {https://doi.org/10.1103/PhysRevD.19.473} {\bibfield
  {journal} {\bibinfo  {journal} {Physical Review D}\ }\textbf {\bibinfo
  {volume} {19}},\ \bibinfo {pages} {473} (\bibinfo {year} {1979})}\BibitemShut
  {NoStop}%
\bibitem [{\citenamefont {Greenberger}\ and\ \citenamefont
  {Yasin}(1988)}]{greenberger1988}%
  \BibitemOpen
  \bibfield  {author} {\bibinfo {author} {\bibfnamefont {D.~M.}\ \bibnamefont
  {Greenberger}}\ and\ \bibinfo {author} {\bibfnamefont {A.}~\bibnamefont
  {Yasin}},\ }\href {https://doi.org/10.1016/0375-9601(88)90114-4} {\bibfield
  {journal} {\bibinfo  {journal} {Physics Letters A}\ }\textbf {\bibinfo
  {volume} {128}},\ \bibinfo {pages} {391} (\bibinfo {year}
  {1988})}\BibitemShut {NoStop}%
\bibitem [{\citenamefont {Jaeger}\ \emph {et~al.}(1995)\citenamefont {Jaeger},
  \citenamefont {Shimony},\ and\ \citenamefont {Vaidman}}]{jaeger1995}%
  \BibitemOpen
  \bibfield  {author} {\bibinfo {author} {\bibfnamefont {G.}~\bibnamefont
  {Jaeger}}, \bibinfo {author} {\bibfnamefont {A.}~\bibnamefont {Shimony}},\
  and\ \bibinfo {author} {\bibfnamefont {L.}~\bibnamefont {Vaidman}},\ }\href
  {https://doi.org/10.1103/PhysRevA.51.54} {\bibfield  {journal} {\bibinfo
  {journal} {Physical Review A}\ }\textbf {\bibinfo {volume} {51}},\ \bibinfo
  {pages} {54} (\bibinfo {year} {1995})}\BibitemShut {NoStop}%
\bibitem [{\citenamefont {Englert}(1996)}]{englert1996}%
  \BibitemOpen
  \bibfield  {author} {\bibinfo {author} {\bibfnamefont {B.-G.}\ \bibnamefont
  {Englert}},\ }\href {https://doi.org/10.1103/PhysRevLett.77.2154} {\bibfield
  {journal} {\bibinfo  {journal} {Physical Review Letters}\ }\textbf {\bibinfo
  {volume} {77}},\ \bibinfo {pages} {2154} (\bibinfo {year}
  {1996})}\BibitemShut {NoStop}%
\bibitem [{\citenamefont {Bera}\ \emph {et~al.}(2015)\citenamefont {Bera},
  \citenamefont {Qureshi}, \citenamefont {Siddiqui},\ and\ \citenamefont
  {Pati}}]{bera2015}%
  \BibitemOpen
  \bibfield  {author} {\bibinfo {author} {\bibfnamefont {M.~N.}\ \bibnamefont
  {Bera}}, \bibinfo {author} {\bibfnamefont {T.}~\bibnamefont {Qureshi}},
  \bibinfo {author} {\bibfnamefont {M.~A.}\ \bibnamefont {Siddiqui}},\ and\
  \bibinfo {author} {\bibfnamefont {A.~K.}\ \bibnamefont {Pati}},\ }\href
  {https://doi.org/10.1103/PhysRevA.92.012118} {\bibfield  {journal} {\bibinfo
  {journal} {Physical Review A}\ }\textbf {\bibinfo {volume} {92}},\ \bibinfo
  {pages} {012118} (\bibinfo {year} {2015})}\BibitemShut {NoStop}%
\bibitem [{\citenamefont {Gardiner}\ and\ \citenamefont
  {Collett}(1985)}]{gardiner1985c}%
  \BibitemOpen
  \bibfield  {author} {\bibinfo {author} {\bibfnamefont {C.~W.}\ \bibnamefont
  {Gardiner}}\ and\ \bibinfo {author} {\bibfnamefont {M.~J.}\ \bibnamefont
  {Collett}},\ }\href {https://doi.org/10.1103/PhysRevA.31.3761} {\bibfield
  {journal} {\bibinfo  {journal} {Physical Review A}\ }\textbf {\bibinfo
  {volume} {31}},\ \bibinfo {pages} {3761} (\bibinfo {year}
  {1985})}\BibitemShut {NoStop}%
\bibitem [{\citenamefont {Gardiner}\ and\ \citenamefont
  {Zoller}(2010)}]{gardiner2010a}%
  \BibitemOpen
  \bibfield  {author} {\bibinfo {author} {\bibfnamefont {C.~W.}\ \bibnamefont
  {Gardiner}}\ and\ \bibinfo {author} {\bibfnamefont {P.}~\bibnamefont
  {Zoller}},\ }\href@noop {} {\emph {\bibinfo {title} {Quantum Noise: A
  Handbook of {{Markovian}} and Non-{{Markovian}} Quantum Stochastic Methods
  with Applications to Quantum Optics}}},\ \bibinfo {edition} {3rd}\ ed.,\
  Springer Series in Synergetics\ (\bibinfo  {publisher} {Springer},\ \bibinfo
  {address} {Berlin Heidelberg},\ \bibinfo {year} {2010})\BibitemShut {NoStop}%
\bibitem [{\citenamefont {Walls}\ and\ \citenamefont
  {Milburn}(2008)}]{walls2008}%
  \BibitemOpen
  \bibfield  {author} {\bibinfo {author} {\bibfnamefont {D.~F.}\ \bibnamefont
  {Walls}}\ and\ \bibinfo {author} {\bibfnamefont {G.~J.}\ \bibnamefont
  {Milburn}},\ }\href {https://doi.org/10.1007/978-3-540-28574-8} {\emph
  {\bibinfo {title} {Quantum {{Optics}}}}},\ \bibinfo {edition} {2nd}\ ed.,\
  Springer {{eBook Collection}}\ (\bibinfo  {publisher} {Springer Berlin
  Heidelberg},\ \bibinfo {address} {Berlin, Heidelberg},\ \bibinfo {year}
  {2008})\BibitemShut {NoStop}%
\bibitem [{\citenamefont {Blais}\ \emph {et~al.}(2004)\citenamefont {Blais},
  \citenamefont {Huang}, \citenamefont {Wallraff}, \citenamefont {Girvin},\
  and\ \citenamefont {Schoelkopf}}]{blais2004}%
  \BibitemOpen
  \bibfield  {author} {\bibinfo {author} {\bibfnamefont {A.}~\bibnamefont
  {Blais}}, \bibinfo {author} {\bibfnamefont {R.-S.}\ \bibnamefont {Huang}},
  \bibinfo {author} {\bibfnamefont {A.}~\bibnamefont {Wallraff}}, \bibinfo
  {author} {\bibfnamefont {S.~M.}\ \bibnamefont {Girvin}},\ and\ \bibinfo
  {author} {\bibfnamefont {R.~J.}\ \bibnamefont {Schoelkopf}},\ }\href
  {https://doi.org/10.1103/PhysRevA.69.062320} {\bibfield  {journal} {\bibinfo
  {journal} {Physical Review A}\ }\textbf {\bibinfo {volume} {69}},\ \bibinfo
  {pages} {062320} (\bibinfo {year} {2004})}\BibitemShut {NoStop}%
\bibitem [{\citenamefont {Wallraff}\ \emph {et~al.}(2004)\citenamefont
  {Wallraff}, \citenamefont {Schuster}, \citenamefont {Blais}, \citenamefont
  {Frunzio}, \citenamefont {Huang}, \citenamefont {Majer}, \citenamefont
  {Kumar}, \citenamefont {Girvin},\ and\ \citenamefont
  {Schoelkopf}}]{wallraff2004}%
  \BibitemOpen
  \bibfield  {author} {\bibinfo {author} {\bibfnamefont {A.}~\bibnamefont
  {Wallraff}}, \bibinfo {author} {\bibfnamefont {D.~I.}\ \bibnamefont
  {Schuster}}, \bibinfo {author} {\bibfnamefont {A.}~\bibnamefont {Blais}},
  \bibinfo {author} {\bibfnamefont {L.}~\bibnamefont {Frunzio}}, \bibinfo
  {author} {\bibfnamefont {R.-S.}\ \bibnamefont {Huang}}, \bibinfo {author}
  {\bibfnamefont {J.}~\bibnamefont {Majer}}, \bibinfo {author} {\bibfnamefont
  {S.}~\bibnamefont {Kumar}}, \bibinfo {author} {\bibfnamefont {S.~M.}\
  \bibnamefont {Girvin}},\ and\ \bibinfo {author} {\bibfnamefont {R.~J.}\
  \bibnamefont {Schoelkopf}},\ }\href {https://doi.org/10.1038/nature02851}
  {\bibfield  {journal} {\bibinfo  {journal} {Nature}\ }\textbf {\bibinfo
  {volume} {431}},\ \bibinfo {pages} {162} (\bibinfo {year}
  {2004})}\BibitemShut {NoStop}%
\bibitem [{\citenamefont {Aspelmeyer}\ \emph {et~al.}(2014)\citenamefont
  {Aspelmeyer}, \citenamefont {Kippenberg},\ and\ \citenamefont
  {Marquardt}}]{aspelmeyer2014a}%
  \BibitemOpen
  \bibfield  {author} {\bibinfo {author} {\bibfnamefont {M.}~\bibnamefont
  {Aspelmeyer}}, \bibinfo {author} {\bibfnamefont {T.~J.}\ \bibnamefont
  {Kippenberg}},\ and\ \bibinfo {author} {\bibfnamefont {F.}~\bibnamefont
  {Marquardt}},\ }\href {https://doi.org/10.1103/RevModPhys.86.1391} {\bibfield
   {journal} {\bibinfo  {journal} {Reviews of Modern Physics}\ }\textbf
  {\bibinfo {volume} {86}},\ \bibinfo {pages} {1391} (\bibinfo {year}
  {2014})}\BibitemShut {NoStop}%
\bibitem [{\citenamefont {Holzman}\ and\ \citenamefont
  {Ivry}(2019)}]{holzman2019}%
  \BibitemOpen
  \bibfield  {author} {\bibinfo {author} {\bibfnamefont {I.}~\bibnamefont
  {Holzman}}\ and\ \bibinfo {author} {\bibfnamefont {Y.}~\bibnamefont {Ivry}},\
  }\href {https://doi.org/10.1002/qute.201800058} {\bibfield  {journal}
  {\bibinfo  {journal} {Advanced Quantum Technologies}\ }\textbf {\bibinfo
  {volume} {2}},\ \bibinfo {pages} {1800058} (\bibinfo {year}
  {2019})}\BibitemShut {NoStop}%
\bibitem [{\citenamefont {Galiffi}\ \emph {et~al.}(2022)\citenamefont
  {Galiffi}, \citenamefont {Tirole}, \citenamefont {Yin}, \citenamefont {Li},
  \citenamefont {Vezzoli}, \citenamefont {Huidobro}, \citenamefont
  {Silveirinha}, \citenamefont {Sapienza}, \citenamefont {Al{\`u}},\ and\
  \citenamefont {Pendry}}]{galiffi2022}%
  \BibitemOpen
  \bibfield  {author} {\bibinfo {author} {\bibfnamefont {E.}~\bibnamefont
  {Galiffi}}, \bibinfo {author} {\bibfnamefont {R.}~\bibnamefont {Tirole}},
  \bibinfo {author} {\bibfnamefont {S.}~\bibnamefont {Yin}}, \bibinfo {author}
  {\bibfnamefont {H.}~\bibnamefont {Li}}, \bibinfo {author} {\bibfnamefont
  {S.}~\bibnamefont {Vezzoli}}, \bibinfo {author} {\bibfnamefont {P.~A.}\
  \bibnamefont {Huidobro}}, \bibinfo {author} {\bibfnamefont {M.~G.}\
  \bibnamefont {Silveirinha}}, \bibinfo {author} {\bibfnamefont
  {R.}~\bibnamefont {Sapienza}}, \bibinfo {author} {\bibfnamefont
  {A.}~\bibnamefont {Al{\`u}}},\ and\ \bibinfo {author} {\bibfnamefont {J.~B.}\
  \bibnamefont {Pendry}},\ }\bibfield  {journal} {\bibinfo  {journal} {Advanced
  Photonics}\ }\textbf {\bibinfo {volume} {4}},\ \href
  {https://doi.org/10.1117/1.AP.4.1.014002} {10.1117/1.AP.4.1.014002} (\bibinfo
  {year} {2022})\BibitemShut {NoStop}%
\bibitem [{\citenamefont {Harwood}\ \emph
  {et~al.}(2025{\natexlab{b}})\citenamefont {Harwood}, \citenamefont {Vezzoli},
  \citenamefont {Raziman}, \citenamefont {Hooper}, \citenamefont {Tirole},
  \citenamefont {Wu}, \citenamefont {Maier}, \citenamefont {Pendry},
  \citenamefont {Horsley},\ and\ \citenamefont {Sapienza}}]{harwood2025a}%
  \BibitemOpen
  \bibfield  {author} {\bibinfo {author} {\bibfnamefont {A.~C.}\ \bibnamefont
  {Harwood}}, \bibinfo {author} {\bibfnamefont {S.}~\bibnamefont {Vezzoli}},
  \bibinfo {author} {\bibfnamefont {T.~V.}\ \bibnamefont {Raziman}}, \bibinfo
  {author} {\bibfnamefont {C.}~\bibnamefont {Hooper}}, \bibinfo {author}
  {\bibfnamefont {R.}~\bibnamefont {Tirole}}, \bibinfo {author} {\bibfnamefont
  {F.}~\bibnamefont {Wu}}, \bibinfo {author} {\bibfnamefont {S.~A.}\
  \bibnamefont {Maier}}, \bibinfo {author} {\bibfnamefont {J.~B.}\ \bibnamefont
  {Pendry}}, \bibinfo {author} {\bibfnamefont {S.~A.~R.}\ \bibnamefont
  {Horsley}},\ and\ \bibinfo {author} {\bibfnamefont {R.}~\bibnamefont
  {Sapienza}},\ }\href {https://doi.org/10.1038/s41467-025-60159-9} {\bibfield
  {journal} {\bibinfo  {journal} {Nature Communications}\ }\textbf {\bibinfo
  {volume} {16}},\ \bibinfo {pages} {5147} (\bibinfo {year}
  {2025}{\natexlab{b}})}\BibitemShut {NoStop}%
\end{thebibliography}
%

%
%
%

\end{document}